\documentclass[aps,prl,twocolumn]{revtex4-1}
\usepackage{amssymb}
\usepackage{bm}
\usepackage{float}
\usepackage{graphicx}
\begin{document}
\title{Quantum tunneling of a vortex between two pinning potentials}
\author{O. Fialko$^1$}
\author{A. S. Bradley$^2$}
\author{J. Brand$^1$}
\affiliation{$^1$Centre for Theoretical Chemistry and Physics, New Zealand Institute for Advanced Study, 
Massey University (Albany Campus), Auckland, New Zealand \\
$^2$Jack Dodd Centre for Quantum Technology, Department of Physics, University of Otago, Dunedin, New Zealand}
\date{\today}

\begin{abstract}
A vortex can tunnel between two pinning potentials in an atomic Bose-Einstein condensate on a time scale of the order of 1s under typical experimental conditions. This makes it possible to detect the tunneling experimentally.  We calculate the tunneling rate by phenomenologically treating vortices as point-like charged particles moving in an inhomogeneous magnetic field. The obtained results 
are in close agreement with numerical simulations based on the stochastic c-field theory.


\end{abstract}

\maketitle
Topologically stable quantized vorticity is the smoking gun of superfluidity and has been 
studied extensively
in superconductors, 
superfluid Helium and most recently in Bose-Einstein condensed (BEC) ultra-cold atomic gases  \cite{leggett}.
Tuneability of experimental parameters in dilute-gas systems by means of Feshbach resonances
and external magnetic or electric fields make it possible to investigate fundamental
condensed matter phenomena in different regimes that are not otherwise easily accessible~\cite{lewen07,bloch08}.
Recent experimental advances in the field include demonstration of quantum phase transitions \cite{bloch05}, the Josephson effect \cite{levy07} and
vortex formation through synthetic gauge fields \cite{Lin09}.

A superfluid vortex in a large BEC is a macroscopic object involving many degrees of freedom and is therefore expected to obey the laws of classical physics. The static and dynamic vortex properties in a BEC have been successfully described by 
the mean-field Gross-Pitaevskii equation \cite{pethick}, which is a classical field theory. If the number of vortices is so large that their 
density is comparable to the particle density,
quantum fluctuations dominate and
the mean-field description fails.
In this case a quantum phase transition  is predicted to take place from the BEC ground state to a highly correlated state,
similar to the Laughlin state in a quantum Hall liquid \cite{fetter10}.
Quantum corrections are also needed if the dynamics of a single vortex over distances comparable to its core size (the healing length) are considered.

Quantum tunneling of particles through a barrier, a counterintuitive consequence of the superposition principle, is known to happen on atomic scales. As pointed out by Schr\"odinger, quantum mechanics admits the superposition of macroscopic states as well [9].
%
An intense search
for macroscopic systems that show evidence 
of  the effect
has so far been successful only for strongly-interacting condensed-matter systems \cite{leggett80, *martinis88, *ustinov03, *many10}. 
Here, we focus on the quantum tunneling of a single superfluid vortex between two adjacent pinning sites in a weakly-interacting BEC. 
This process involves a small, controllable fraction of the total superfluid, thus presenting a route to larger superpositions.
We employ two methods of study. First, we treat the vortex as a charged point particle in an effective inhomogeneous magnetic field,
exhibited by the pinning potentials. We show that a charged particle
feels a double well potential in such a field. We then calculate the tunneling rate in the double well potential 
within the Heitler-London approximation. Second, we simulate the vortex dynamics within the truncated Wigner approximation, 
which takes into account quantum noise on top of the mean-field Gross-Pitaevskii equation. 
The two theories give closely comparable predictions for the tunneling rate, and a much larger rate than found in previous 
studies of vortex tunneling \cite{perelomov71,*auerbach06}.  

A dilute-gas BEC containing a vortex consists of a superfluid whirl around a hole in the condensate. 
In a homogeneous system, vortices located at different positions are degenerate.
In the presence of a localized impurity, the degeneracy is lifted. An impurity that repels atoms will attract and thus pin the vortex. A vortex can even be pinned by a localized impurity in the presence of harmonic trapping as studied in detail in Ref.\ \cite{anderson09}. 

\begin{figure}
\begin{center}
\includegraphics[width=7.5cm,height=5.5cm]{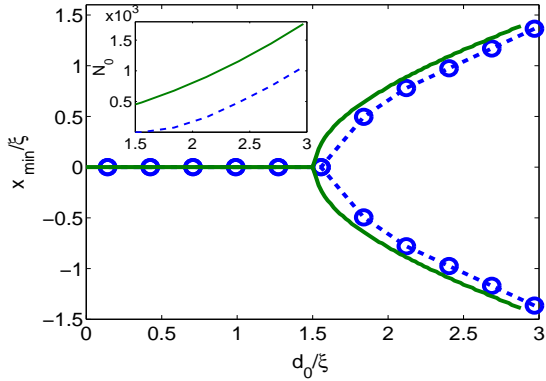}
\end{center}
\caption{Equilibrium positions $x_{\rm min}$ of a vortex in the presence of two pinning potentials separated by the distance $d_0 = 2 r_0$. 
Solid line: analytical result based on Eq. (\ref{naiveDE}). Open circles: 
results of numerical simulations based on the solutions of Eq. (\ref{Eq.GP}). Inset: number of particles inside 
a circle with the radius equal to a pinned position of the vortex. Dashed line: result of the numerical simulation. 
Solid line: analytical result based of the assumption that the vortex is structureless.}
\label{NaiveB}
\end{figure}

Here we study a vortex in the presence of two equivalent pinning potentials. First, we employ mean-field theory to study the possibility
of vortex pinning and postpone the consideration of quantum fluctuations.
We assume that the pinning potentials are located at ${\bf{r}}=\pm{\bf{r}}_0$ and the energy of the condensate is lifted 
by $\lambda [n({\bf r}_0)+n(-{\bf r}_0)]$, where $\lambda>0$ is the strength of the pinning potentials and $n({\bf r})$ is the density of the condensate  
in the absence of the pinning potentials. We approximate the vortex located at ${\bf r}=0$ by the mean-field wave function
$\psi({\bf r})=\sqrt{n_0}r\exp(i\theta)/\sqrt{2\xi^2+r^2}$ \cite{pethick},
such that $n({\bf r})=|\psi({\bf r})|^2$, $\xi$ is the 
healing length and $\theta$ is the polar angle of the vector ${\bf r}$.  
If the core of a vortex is located at ${\bf r}=x\hat{{\bf r}}_0$, then the extra energy due to the pinning potentials is
\begin{eqnarray}
\Delta E(x)\sim\lambda n_0\left[\frac{(x-r_0)^2}{2\xi^2+(x-r_0)^2} + \frac{(x+r_0)^2}{2\xi^2+(x+r_0)^2}\right].\;\;\;
\label{naiveDE}
\end{eqnarray}
We plot $x_{{\rm min}}$ such that $\Delta E(x_{\rm min})=\min_x [\Delta E(x)]$ in Fig.~\ref{NaiveB}. 
We see that if the separation between pins is larger than roughly
twice the healing length, then the vortex can be pinned by one of the pins. On the other hand, if the separation is smaller than that, the vortex is
located between the two pins. This bifurcation is caused by the appearance of an energy barrier between the two pins for the vortex to overcome.

To compare this  with more quantitative results,
we numerically solve the Gross-Pitaevskii
equation in 2D \cite{pethick}:
\begin{eqnarray}
i\hbar\frac{\partial\psi({\bf r},t)}{\partial t}&=&\left(-\frac{\hbar^2{\bf \nabla}^2}{2m} + V({\bf r}) 
+g|\psi({\bf r},t)|^2\right)\psi({\bf r},t).\;\;\;
\label{Eq.GP}
\end{eqnarray}
Here $V({\bf r})=V^1_{\rm pin}({\bf r})+V^2_{\rm pin}({\bf r})+V_{\rm trap}({\bf r})$ consists of three contributions.  
The two pinning potentials are of the form 
$V^{1,2}_{\rm pin}({\bf r})=V_0/(\pi {\Delta}^2) \exp[-({\bf r}\pm {\bf r}_0)^2/\Delta^2]$ with $V_0$ being the strength of a pinning potential, 
$\Delta$ is its width. The trapping potential is $V_{\rm trap}({\bf r})=m\omega_{\bot}^2 {\bf r}^2/2$, where ${\bf r}$ is the 2D vector. The  two-body interaction strength in 2D 
$g=\sqrt{8\pi}\hbar^2a/(ma_z)$ is 
expressed via the s-wave scattering length 
$a$ and the axial oscillator length 
$a_{z}=\sqrt{\hbar/m\omega_{z}}$. The axial confinement 
is assumed much stronger than the radial one
$a_{\bot}=\sqrt{\hbar/m\omega_{\bot}}\gg a_{z}$.
We have solved Eq. (\ref{Eq.GP}) for a stationary state with a pinned vortex. The healing length is defined as $\xi=\hbar/\sqrt{2mgn_p}$,
where $n_p$ is the local density of the condensate at a pinning potential in the absence of the vortex. We have chosen
$\Delta = 0.1a_{\perp}$, $g=0.01\hbar^2/m$, $V_0=4\hbar^2/m$ and the number of particles
$N=\int d{\bf r}|\psi({\bf r})|^2\approx 2\times 10^5$. 
This corresponds to the chemical potential $\mu=25\hbar\omega_{\bot}$ and the pinning potential amplitude  $V_0/ (\pi {\Delta}^2) = 5\mu$.
We find that the vortex can be pinned by one of the pins if the separation between them is larger than a critical value in agreement with the
previous result. Both are shown in Fig.\ \ref{NaiveB}.

In the recent experiment with $^{23}$Na atoms \cite{ryu09}, $\omega_z/2\pi=1$kHz and $\omega_{\bot}/2\pi=20$Hz.
This gives $a_{\perp}\approx 4.7\mu m$, 
$\xi\approx 1.6\mu m$ and the bulk density $n_0\approx 83\mu m^{-2}\approx 5.7 n_p$.
Narrow pinning potentials could be realized by blue-detuned lasers focused to a beam of diameter of  $\sim 0.7\mu m$ at half maximum or by heavy atoms in a species-specific optical potential of double-well shape \cite{Helmerson}. 
The vortex can be created by standard techniques~\cite{dalibard00} and located on one specific pin by adiabatic ramping procedures. After a given hold time, the vortex location can be measured by imaging after a brief expansion phase without pinning potentials. This would reveal experimentally whether tunneling has occurred.

A vortex 
in a density gradient
feels a Magnus force, which is analogous to the Lorentz force on a charged particle in a magnetic field. Here, we further develop a phenomenological approach based on this analogy \cite{thouless93,thouless94} in order to estimate the rate of vortex tunneling between the pins.
We approximate the vortex located at ${\bf r}$ 
by the Feynman many-body  wave-function \cite{feynman56}
\begin{eqnarray}
\langle{\bf r}_1,  \ldots, {\bf r}_N |\psi\rangle&=&\prod_{j=1}^{N}e^{i\theta({\bf r}_j-{\bf r})}\psi_0({\bf r}_1,..., {\bf r}_N; {\bf r}).
\label{Eq.Feynman}
\end{eqnarray}
Here $\theta({\bf r}_j-{\bf r})=\arctan[(y_j-y)/(x_j-x)]$ 
is the phase of the atom at ${\bf r}_j$ relative to the center of the vortex at ${\bf r}$. 
The real function $\psi_0$ vanishes at the vortex core and approaches $\sqrt{n}$ far away from it. When the vortex moves,
the value ${\bf r}$ changes and the vortex state acquires a phase factor. This is similar to the Aharonov-Bohm effect 
\cite{bohm59}
and motivates the analogy
with a charged particle.

\begin{figure}[t]
\begin{center}
\includegraphics[width=7.5cm,height=5cm]{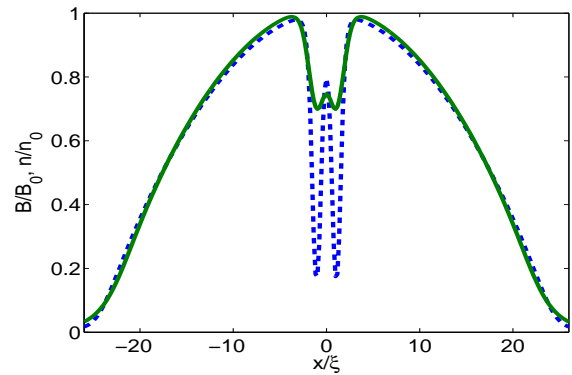}
\end{center}
\caption{Effective magnetic field $B$ along the axis through the trap center (solid line) and  density $n$ (dashed line). 
The two pinning potentials are separated by the distance  $d_0=2\xi$.
$B_0=n_0 h/q$, where $n_0$ is the bulk density at the trap center.  
In this configuration the vortex is pinned classically
(cf.\ Fig.\ \ref{NaiveB}) and a 
double-well potential 
appears 
within the effective theory.}
\label{magn1p}
\end{figure}

A 
particle with charge $q$ moving in a magnetic field  ${\bf A}({\bf r})$ along a path $\Gamma$ 
acquires an Aharonov-Bohm phase $\Delta\theta=-\frac{q}{\hbar}\int_{\Gamma}
{\bf A}({\bf r})d{\bf r}$, such that its state vector changes as $|\psi\rangle \rightarrow e^{-i\Delta\theta}|\psi\rangle $.
This can also be calculated as
$\Delta\theta=-{\rm Im}\int_{\Gamma}d{\bf r}\langle\psi|{\bf \nabla}_{{\bf r}}\psi\rangle$. When we substitute into this expression 
the Feynman variational state from Eq.~(\ref{Eq.Feynman}), we obtain
$\Delta\theta_v=-\int d^2{r}'\int_{\Gamma}d{\bf r}[{\bf \nabla}_{{\bf r}} \theta({\bf r}'-{\bf r})]\rho({\bf r}';{\bf r})$.
Here $\rho({\bf r}';{\bf r})=N\int d^2 {r}_2...d^2{r}_N\psi_0^2({\bf r}',{\bf r}_2,..., {\bf r}_{N}; {\bf r})$ 
is the particle density at ${\bf r}'$ of the BEC  with a vortex core at ${\bf r}$. 
Comparing the expressions $\Delta\theta$ and $\Delta\theta_v$ for the phase change,
we conclude that
we may regard a vortex as a charged particle 
subject to a vector potential
\begin{eqnarray}
{\bf A}({\bf r})&=&\frac{\hbar}{q}\int d^2{\bf r}'[{\bf \nabla}_{{\bf r}} \theta({\bf r}'-{\bf r})]\rho({\bf r}';{\bf r}).
\label{EqA}
\end{eqnarray}
We approximate the density 
of a vortex at ${\bf r}$ by    
$\rho({\bf r}';{\bf r})=n({\bf r}') |{\bf r}'-{\bf r}|^2/[2\xi^2({\bf r}')+|{\bf r}'-{\bf r}|^2]$ \cite{pethick},
where $n({\bf r}')$ is the mean-field density without a vortex from Eq.~(\ref{Eq.GP}) and $\xi({\bf r}')=\hbar/\sqrt{2mgn({\bf r}')}$ the local healing length.
From Eq.~(\ref{EqA}) we calculate the magnetic field ${\bf B}={\bf \nabla}\times{\bf A}$.
It is directed along the $z$ axis.
The magnitude is proportional to the convoluted density, blurred on the size scale of the vortex core, as see in Fig.~\ref{magn1p}.
We shall study the dynamics of the charged particle in this inhomogeneous  
magnetic field.

A quantum particle with charge $q$ in a constant magnetic field ${\bf B}=B{\bf e}_z$ and a square box of size $L\times L$
has energy levels $E_n=\hbar\omega_c(n+1/2)$, where $\omega_c=qB/m$ \cite{schwabl}.
Each level is highly degenerate. The lowest Landau level is characterized by the wave functions 
\begin{equation}
\langle {\bf r} | {\bf r}_q\rangle=
\frac{1}{\sqrt{2\pi l^2}}\exp{\left(-\frac{|{\bf r}-{\bf r}_q|^2}{4l^2}+\frac{i{\bf e}_z\cdot [{\bf r}_q \times {\bf r}]}{2l^2}\right)},
\label{EqLandau}
\end{equation}
with $l^2=\hbar/(qB)$.
Degeneracy corresponds to different ${\bf r}_q$, which are separated by $|\Delta {\bf r}_q|=l^2/L$. 
Thus, the lowest state will be a superposition of many such wave functions
with different ${\bf r}_q$. 
The particle can be found at any ${\bf r}_q$ with equal probability. 

It is difficult to deal with this degeneracy numerically. A simplified treatment with an effective Hamiltonian can be found when the Landau length $l$ is much smaller than the healing length $\xi$. Formally, we assume delocalized Landau wave functions on spatial domains ${\cal R}$ of the size $\sim \xi^2$ where we approximate $B({\bf r})$ as constant, but account for the energy dependence of the lowest Landau level on larger length scales. Averaging over the domains ${\cal R}$ then yields the effective Hamiltonian
\begin{equation}
\hat{H}_{\rm eff}= \frac{\hat{\bf p}^2}{2m}+\frac{\hbar q B({\bf r})}{2m}.
\label{Heff}
\end{equation}
Intuitively, this can be understood as a Zeeman Hamiltonian for a particle with orbital magnetic moment $\mu_o = \hbar q/(2m)$, 
which is obtained semiclassically for 
a charged particle orbiting at a radius given by the Landau length $l$, with cyclotron frequency $\omega_c$.

The magnetic field depicted in Fig.\ \ref{magn1p} suggests that Eq. (\ref{Heff}) describes a particle moving in a double well potential.
The tunneling rate of the particle between two wells gives the tunneling rate of the vortex between two pins.

\begin{figure}[t]
\begin{center}
\includegraphics[width=7.5cm,height=5cm]{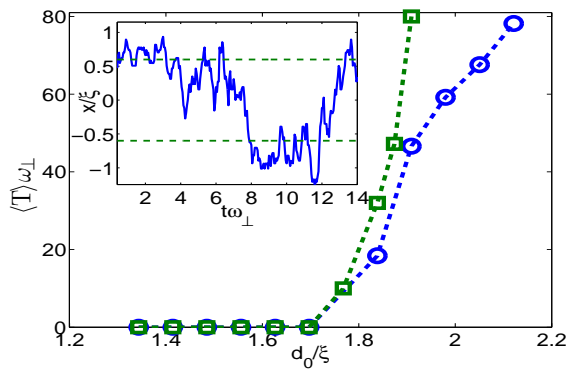}
\end{center}
\caption{The period of quantum oscillations of a vortex between two pining potentials as a function of distance between them. 
The open circles represent the result of the truncated Wigner approximation. The open squares are the result of 
the effective theory (see text). Inset: One of the trajectories of a vortex. Dashed lines indicate the $x_{\bf min}\approx\pm 0.6\xi$ 
values corresponding to $d_0\approx 1.8\xi$ in Fig.\ \ref{NaiveB}.}
\label{Fig.period}
\end{figure}

The vortex tunneling rate can be related to a wave-function overlap within the Heitler-London approximation \cite{mattis, auerbach06}. 
Let $|\psi_{i}\rangle$ ($i=1,2$) be coherent states of the vortex pinned by one of the pinning potentials.
We construct symmetric and antisymmetric states as
$|\psi_{\pm}\rangle=(|\psi_1\rangle\pm |\psi_2\rangle)/\sqrt{2(1\pm|\langle\psi_1|\psi_2\rangle|)}$.
The tunneling rate is then given by
$t_{v}\approx \langle\psi_2|V^1_{\rm pin}|\psi_2\rangle |\langle\psi_1|\psi_2\rangle|$.
The quantity $\langle\psi_2|V^1_{\rm pin}|\psi_2\rangle$
is approximately equal to $V_0 n_p$.  
The period of the coherent oscillations is $T=2\pi\hbar/t_{v}$ and thus
\begin{eqnarray}
T&\approx& \frac{2\pi\hbar}{V_0 n_p |\langle\psi_1|\psi_2\rangle|}.
\label{Eq.heitler}
\end{eqnarray}
The states $|\psi_{i}\rangle$ are the solutions of Eq.~(\ref{Heff}) with only the left and the right pin respectively. 
The tunneling rate shown by circles in Fig.~\ref{Fig.period} was
calculated from Eq.~(\ref{Eq.heitler}) based on the
mean-field density 
and numerical solutions of  Eq.~(\ref{Heff}).
In the following we will 
compare this result with simulations in the truncated Wigner approximation.

Before proceeding, we briefly analyze an approach for calculating the tunneling rate of a vortex found in previous studies. 
In Ref.\ \cite{perelomov71,thouless94} it was assumed that the vortex is a point-like object, i.e.\ the detailed density structure was not taken into
account. This amounts to considering the effective theory developed in the previous section with a constant magnetic field.
Indeed, the structureless vortex has a healing length which is zero and thus
$\rho({\bf r}',{\bf r})=n_0$ in Eq.~(\ref{EqA}). 
In this case
\begin{eqnarray}
{\bf A}({\bf r})=n_0\frac{\hbar}{q}\int d^2{\bf r}'[{\bf \nabla}_{{\bf r}} \theta({\bf r}'-{\bf r})]=
\frac{n_0 h}{2q}({\bf e}_x y-{\bf e}_y x).\;\;\;
\end{eqnarray}
This corresponds to the constant magnetic 
field pointing in the third direction ${\bf B}=n_0 h/q{\bf e}_z$.
The coherent state of a charged particle moving in this magnetic field is the lowest Landau level centered at arbitrary  ${\bf r}_q$ and given
by Eq.~(\ref{EqLandau}).
It is assumed that this degeneracy is broken by the pinning potentials and thus ${\bf r}_q$ is the position of one of them.
Then the overlap of the two states $|{\bf r}_q\rangle$ and $|-{\bf r}_q\rangle$ is
$|\langle -{\bf r}_q|{\bf r}_q\rangle| = \exp{(-2\pi r_q^2n_0)}$.
The distance between two pinning potentials should be of the order of $l$ to ensure that the overlap has a reasonably large
value. On the other hand,  $l$ is of the order of the average inter-particle distance. 
The latter is much smaller than the healing length, i.e.\ the size of the vortex core, which was ignored in the above analysis. 
Indeed, we have $\xi/l\approx 18$. We found that the overlap is 
$\sim\exp{(-N_0)}$, where $N_0 \sim 10^{-3}$ for separations above the bifurcation point in Fig.~\ref{NaiveB} and thus the overlap is practically zero.
This would give very large values for the period of oscillations in Eq. (\ref{Eq.heitler}). 
In contrast, by accounting for the vortex core structure, we have found reasonable values for the period.
We attribute this to two factors: First, the effective number of particles involved in the tunneling process $N_0$ is  reduced. 
This is seen from the inset in Fig.~\ref{NaiveB} and is due to the reduced density at and between the pinning potentials, 
the finite vortex core size and a small deviation between the pin and vortex positions. 
Second, the double well felt by the vortex becomes shallow for small separation of the pins,  allowing for much larger overlap of the localized states. Thus the exponential behavior
found in the above analysis breaks down.


We perform stochastic simulations in the truncated Wigner representation 
to calculate the tunneling rate numerically. This method enables one to capture many quantum features of the system, being increasingly accurate for short evolution times \cite{Steel1998,*Sinatra2001,*blakie08,*Polkovnikov10a}.
We first find a stationary solution of Eq.~(\ref{Eq.GP}) that corresponds to a pinned state. This is one of the states
for $d_0/\xi$ larger than the bifurcation point in Fig.\ \ref{NaiveB}. 
Quantum noise is added to this state as
$\psi({\bf{r}}) \rightarrow \psi({\bf{r}}) + \sum_{j=1}^M f_j({\bf{r}})\alpha_j$,
where $f_j({\bf{r}})$ is an orthonormal basis. 
We use plane waves, $f_j({\bf{r}})=\exp(i{\bf k}_j{\bf r})/\sqrt{V}$ and
$\alpha_j$ are complex-gaussian random variables with 
$\overline{\alpha_i^{\ast}\alpha_j}=\delta_{ij}/2$, sampling vacuum fluctuations in the Wigner representation. 
$M$ is chosen to represent the physical system, while excluding excess vacuum noise~\cite{Steel1998,*Sinatra2001,*blakie08,*Polkovnikov10a}. The results should not change significantly upon appreciable change of $M$.
Two different energy cut-offs $\hbar^2 k^2_{M}/2m = 300 \hbar\omega_{\perp}$ and $400 \hbar\omega_{\perp}$ are used to determine values for $M$ and define the low-energy c-field region.
We then propagate this state in real time by solving the time-dependent Eq. (\ref{Eq.GP}).
One of the trajectories is presented in Fig.\ \ref{Fig.period}. This corresponds to $d_0\approx 1.8\xi$ in Fig.\ \ref{NaiveB}. 
It is seen that the vortex performs oscillatory motion between the two pins with the period $T\approx 14/\omega_{\perp}$. 
We average the period of such oscillations over 100 trajectories and present the result in Fig.\ \ref{Fig.period}.  Our results are indistinguishable on the scale of the figure 
for the two cut-offs. 
We compare the results of the numerical simulations with the results of the effective theory given by Eq. (\ref{Eq.heitler}) in the same figure. 
The results essentially agree in the vicinity of the bifurcation point.  We see that the period of the vortex oscillations
between two pins can be of the order of  1s  if the separation between pins is of the order of two-three lengths. 

In conclusion, we have studied the quantum dynamics of a vortex in the presence of two pinning potentials.
In particular, we have demonstrated that the vortex may quantum mechanically tunnel between the two pins.
The time scale of the tunneling is achievable in current experiments if the separation between two pins is of the order of a few healing lengths. 
If realized experimentally, this will demonstrate unambiguously the possibility of macroscopic quantum tunneling. 

We thank Mikkel Anderson, Peter Drummond and Kristian Helmerson for useful discussions. OF and JB were supported by the Marsden Fund (contract No.~MAU0910) 
administered by the Royal Society of New Zealand. AB received support from the Marsden Fund and the New Zealand Foundation for Research, 
Science, and Technology (contracts UOOX0801 and NERF-UOOX0703).
\bibliographystyle{prsty}


\end{document}